# AN OVERVIEW OF WIRELESS SENSOR NETWORK SECURITY ATTACKS: MODE OF OPERATION, SEVERITY AND MITIGATION TECHNIQUES.

Onwuegbuzie Innocent Uzougbo[1], Samuel-Soma M. Ajibade[2], Fele Taiwo[3]
innoslov@gmail.com[1], samuel.soma@yahoo.com[2], tayewofele@gmail.com[3]

Department of Computer Science
The Federal Polytechnic Ado Ekiti, Ekiti State, Nigeria

## ABSTRACT

Wireless Sensor Network (WSN) is the network of the future. As it gradually gains ground by transforming our lives and environments into a Smart World, it will definitely call for attention from selfish minded Attackers. The first section of this paper introduces the Wireless Sensor Network, its constraints, architecture, and mode of operation. It goes ahead to discuss the applications of WSN in health, agriculture, military, transportation, environment, industries, etc. Unfortunately, WSN automatically inherits the security challenges of the traditional network. In this section, the Network security goals which are Confidentiality, Integrity, Availability, and Authentication sometimes called the CIAA of Network security is discussed with respect to WSN. The WSN Network Protocols Stack is followed, unlike the TCP/IP Network Protocol Stack, the WSN Network Protocol Stack is made up of five (5) layers (i.e. Application, Transportation, Network, Data Link, and Physical layers). Emphasis and informed discussion are made on each layer. This is immediately followed by the security challenges of WSN, these challenges are categorized into two (2), Passive and Active Attacks, Passive attack is interested in the message, data or information that traverses the network without hindering the network channel or medium of communication while Active attack is interested in both the data as well as compromising and if possible shutting down the communication channel in order to hinders the smooth running of the network. In line with these security challenges, mitigation techniques are proffered to possibly prevent or lessen the severity of the attacks as its almost impossible to stop attackers from carrying out their malicious activities. The conclusion brings this paper to an end and obligates researcher to help improve the current mitigation techniques as well as develop entirely new and effective mitigation solutions.

# 1. INTRODUCTION

The time has come. Consciously or unconsciously, the world is inevitably advancing toward becoming a Smart World where almost everything communicates to themselves with or without human intervention and even goes further to make decisions and take adequate actions on behalf of humans. Animate and inanimate objects tagged with very small to almost invisible devices are able to collect data from their surrounding and feed them through appropriate channels to locations where these they can be analysed and used to make informed decisions and actions. These days we are beginning to see the emergence of Smart Cars, Smart Homes, Smart Industries, etc. As the world begins to gain her "Smartness", there is a high level of concern on the security of information that will be and already traversing the Wireless Sensor Network. If this concern is not critically and urgently attended to at this early stage, Attacker would find new profession; that is the practice of compromising and shutting down part a or a whole section of a Smart environment thereby crippling activities that could cause the loss of millions of Dollar per unit time.

This paper discusses Wireless Sensor Network(s) (WSNs), which is the backbone, wired-framework or data high way of a Smart environment. Different types of security challenges facing WSN is discussed with respect to their nature and mode of attacks in line with the with goals of network security (i.e. Confidentiality, Integrity, Availability, and Authentication). Efforts on how to mitigate these threats are also discussed and the conclusion is drawn.

# 2. WIRELESS SENSOR NETWORK AN OVERVIEW

A Smart world is created by Smart Cities and Smart Cities are created and anchored on Wireless Sensor Networks (WSN) backbone, then the question arises, what is Wireless Sensor Network? Smart environments are increasingly being deployed in homes, military, health, ecological, industrial, and transportation applications, among others. These environments are normally based on smart devices that acquire data from the real world, process and communicate these data to information processing centers, generating some information-based services and, sometimes, producing some actions in the environment [1]

Wireless Sensor Networks (WSNs) can be defined as a self-configured and infrastructure-less wireless networks to monitor physical or environmental conditions, such as temperature, sound, vibration, pressure, motion or pollutants and to cooperatively pass their data through the network to a main location or sink where the data can be observed and analysed [2].Wireless sensor networks consist of a large number of low-cost, low-power, and multifunctional sensor nodes that communicate over short distances through wireless links [3].

A sink or base station acts like an interface between users and the network. One can retrieve required information from the network by injecting queries and gathering results from the sink. Typically a wireless sensor network contains hundreds of thousands of sensor nodes. The sensor nodes can communicate among themselves using radio signals. A wireless sensor node is equipped with sensing and computing devices, radio transceivers and power components. The individual nodes in a wireless sensor network (WSN) are inherently resource constrained: they have limited processing speed, storage capacity, and communication bandwidth [2]. They are connected to each other through short-range wireless links, used as an infrastructure to forward the collected data, message or report to an authorized user-end over base station [4] mostly called the Sink, Base Station or Gateway. A typical representation of Wireless Sensor Network is depicted in Figure 1 below.

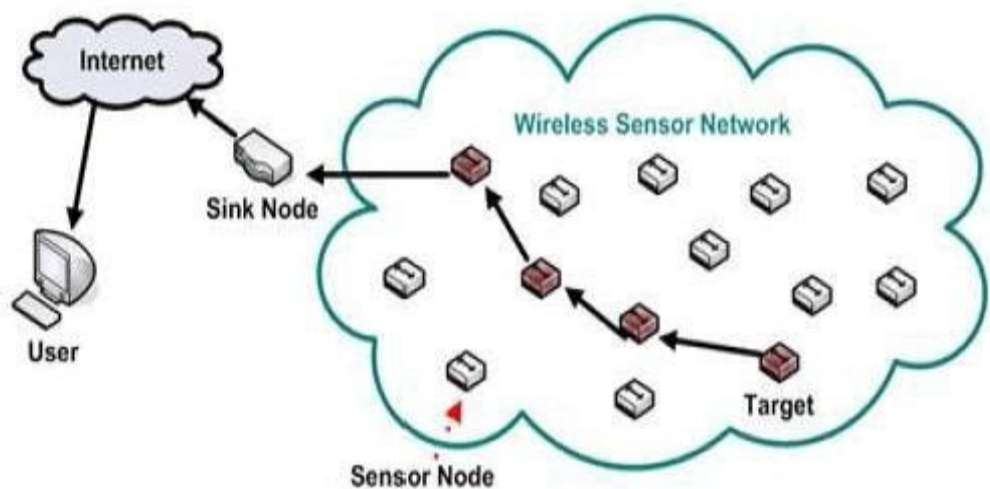

Figure 1: Wireless Sensor Network Architecture.

**3. SENSOR NODE**

The sensor node is the individual device/entity that when configured as a group with respect to a particular routing algorithm forms a wireless network. The sensor node is one of the main parts of a WSN. The hardware of a sensor node generally includes four parts: the power and power management module, a sensor, a micro-controller, and a wireless transceiver [5]
as shown in Figure 2 below. The power module offers the reliable power needed for the system. The sensor, which is a major part of a WSN node can obtain the environmental and equipment status. A sensor is in charge of collecting and transforming the signals, such as light, vibration and chemical signals, into electrical signals and then transferring them to the micro-controller. The micro-controller receives the data from the sensor and processes the data accordingly. The Wireless Transceiver (RF module) then transfers the data, so that the physical realization of communication can be achieved [5]. Figure 2 below depicts the diagram of a typical WSN.

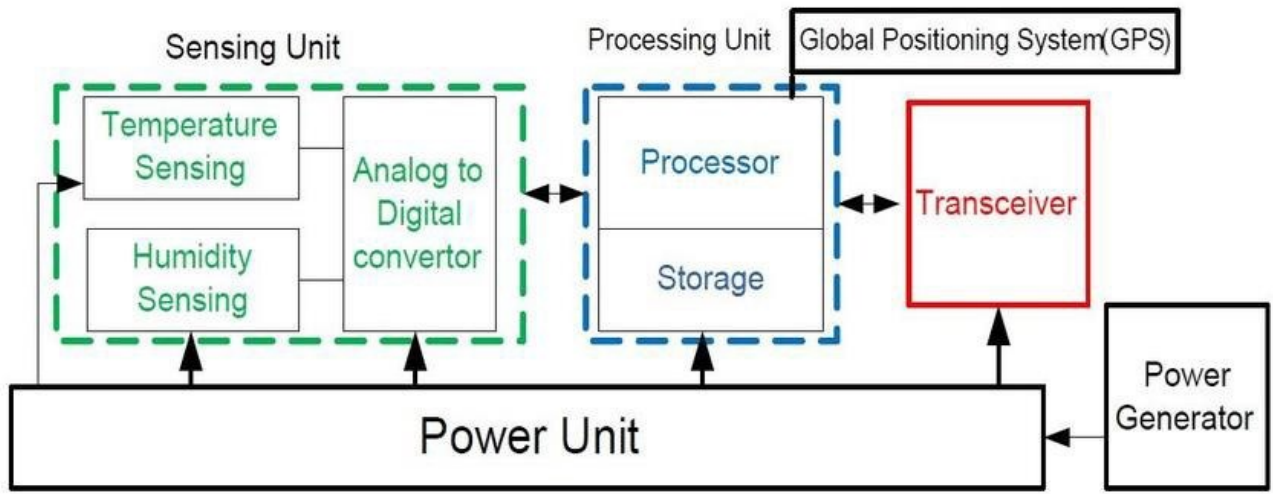

Figure 2: Sensor Node Internal Composition.

WSNs normally employ Wireless Personal Area Network (WPAN) or Low Power Wide Area Network (LPWAN) standards, to relay measured or sensed data to the base station. These standards include IEEE 802.15.4, ZigBee and Bluetooth. There is no single connectivity solution considered suitable for all WSNs, and the choice of standard entirely depends on communication requirements and resource constraints of a particular application [6].

## 4. APPLICATIONS OF WSN

WSN finds its use in various, if not all field of life. In the near future, virtually all things both living and nonliving tagged with a sensor will be connected to a central grid of robust WSN database, where data can be harvested for further processing, management, and decision making.

WSN applications can be classified into two categories: monitoring and tracking. Monitoring applications include indoor or outdoor environmental monitoring, health and wellness monitoring, power monitoring, inventory location monitoring, factory and process automation, and seismic and structural monitoring. Tracking applications include tracking objects, animals, humans, and vehicles [7]. Below are some of its most popular use:

*Military applications:* Gadgets tagged with wireless sensors is an integral part of the military. These smart gadgets assist the military in control, communications, computing, intelligence, battlefield surveillance, reconnaissance and targeting systems.

*Area monitoring:* In area monitoring, the sensor nodes are deployed over a region where some phenomenon is to be monitored. When the sensors detect the event being monitored (heat, pressure etc.), the event is reported to one of the base stations, which then takes appropriate action [2].

*Transportation:* Real-time traffic information is being collected by WSNs to later feed transportation models and alert drivers of congestion and traffic problems.

*Health applications:* Some of the health applications for sensor networks are supporting interfaces for the disabled, integrated patient monitoring, diagnostics, and drug administration in hospitals, telemonitoring of human physiological data, and tracking & monitoring doctors or patients inside a hospital [2].

*Environmental sensing:* The term Environmental Sensor Networks has developed to cover many applications of WSNs to earth science research. This includes sensing volcanoes, oceans, glaciers, forests etc. Some other major areas are listed below:

  i. Air pollution monitoring
  ii. Forest fires detection
  iii. Greenhouse monitoring
  iv. Landslide detection

*Structural monitoring:* Wireless sensors can be utilized to monitor the movement within buildings and infrastructure such as bridges, flyovers, embankments, tunnels etc enabling Engineering practices to monitor assets remotely without the need for costly site visits [2].

*Industrial monitoring:* Wireless sensor networks have been developed for machinery condition-based maintenance (CBM) as they offer significant cost savings and enable new functionalities. In wired systems, the installation of enough sensors is often limited by the cost of wiring.

*Agricultural sector:* using a wireless network frees the farmer from the maintenance of wiring in a difficult environment. Irrigation automation enables more efficient water use and reduces waste [2].

## 5. WIRELESS SENSOR NETWORK SECURITY GOALS AND CHALLENGES

To truly appreciate the severity of network attacks it is important to grasp the various security goals of a network and data traversing or resident on the network. The standard universal security goals of a network are; Confidentiality, Integrity, Availability, and Authenticity, sometimes referred to as the CIAA of network security.

*Confidentiality:* Confidentiality is the assurance that sensitive data is being accessed and viewed only by those who are authorized to see it. But when node replication attack is launched, confidentiality of data is not assured as clone nodes are the duplicated nodes of the compromised ones, and thus they behave like original compromised nodes. These clone nodes can have all the data that contains trade secrets for commercial business, secret classified government information, or private medical or financial records, and thus by misusing such sensitive data, it can damage the network or organization, person, and governmental body [8].

*Integrity:* Integrity ensures that the contents of data or correspondences are preserved and remain unharmed during the transmission from sender to receiver. Integrity represents that there is a guarantee that a message sent is the message received meaning that it was not altered either intentionally or unintentionally during transmission. But in case of node replication attack, an attacker can falsify sensor data or can inject false data to corrupt the sensitive data and thus subverting the data aggregation using the replicated or clone nodes [8].

*Availability:* Availability ensures the survivability of network services despite attacks. In case of node replication attack, an attacker or attacker is able to compromise the availability of WSN by launching a Denial of Service (DoS) attack, which can severely hinder the network's ability to continue its processing. By jamming legitimate signals, the availability of the network assets to authorized parties is also affected [8].

*Authenticity:* Authenticity is a security goal that enables a node to ensure the identity of the sensor node it is communicating with. In case of node replication attack, an attacker creates clone nodes which are seemingly legitimate ones (identical to the originally captured node) as they have all the secret credentials of the captured node; thus, it is difficult for any node to differentiate between a cloned node and the original or legitimate node. Also, the existing authentication techniques cannot detect clone nodes as they all hold legitimate keys. This is how the authenticity of the network is affected [8].

## 6. WSN NETWORK PROTOCOL STACK

Just as there are rules and regulations guiding the proper functioning of all coordinated system so also are rules and regulations guiding the functioning of the Wireless Sensor Network. For two or more Nodes to communicates, lots of layer encapsulations and decapsulation takes place according to the respective layer traversed through by the traversing data, message or information as the case may be. The Network protocol stack of WSN is segmented into five (5) layers as detailed and further accompanied by the diagram in Figure 3 below:

    a)     ***Physical Layer:*** The physical layer is responsible for converting bit streams from the data link layer to signals that are suitable for transmission over the communication medium. For this purpose, it must deal with various related issues, for example, transmission medium and frequency selection, carrier frequency generation, signal modulation and detection, and data encryption. In addition, it must also deal with the design of the underlying hardware, and various electrical and mechanical interfaces.

b)     *Data Link Layer:* The data link layer is responsible for data stream multiplexing, data frame creation and detection, medium access, and error control in order to provide reliable point-to-point and point-to-multipoint transmissions. One of the most important functions of the data link layer is medium access control (MAC). The primary objective of MAC is to fairly and efficiently share the shared communication resources or medium among multiple sensor nodes in order to achieve good network performance in terms of energy consumption, task management, network throughput, and delivery latency.

c)     *Network Layer:* The network layer is responsible for routing the data sensed by source sensor nodes to the data sink(s). In a sensor network, sensor nodes are deployed in a sensing region to observe a phenomenon of interest, or data needed to be transmitted to the data sink. In general, a source node can transmit the sensed data to the sink either directly via single-hop long-range wireless communication or via multi-hop short-range wireless communication. However, long-range wireless communication is costly in terms of both energy consumption and implementation complexity for sensor nodes. In contrast, multi-hop short-range communication can not only significantly reduce the energy consumption of sensor nodes, but also effectively reduce the signal propagation and channel fading effects inherent in long-range wireless communication, and is therefore preferred. Since sensor nodes are densely deployed and neighbor nodes are close to each other, it is possible to use multi-hop short-range communication in sensor networks. In this case, to send the sensed data to the sink, a source node must employ a routing protocol to select an energy efficient multi-hop path from the node itself to the sink.

d)     *Transport Layer:* In general, the transport layer is responsible for reliable end-to-end data delivery between sensor nodes and the sink(s). Due to the energy, computation, and storage constraints of sensor nodes, traditional transport protocols cannot be applied directly to sensor networks without modification. For example, the conventional end-to-end retransmission-based error control and the window-based congestion control mechanisms used in the transport control protocol (TCP) cannot be used for sensor networks directly because they are not efficient in resource utilization.

e)     *Application Layer:* The application layer includes a variety of application-layer protocols that perform various sensor network applications, such as query dissemination, node localization, time synchronization, and network security. For example, the Sensor Management Protocol (SMP) is an application-layer management protocol that provides software operations to perform a variety of tasks, for example, exchanging location-related data, synchronizing sensor nodes, moving sensor nodes, scheduling sensor nodes, and querying the status of sensor nodes [9].

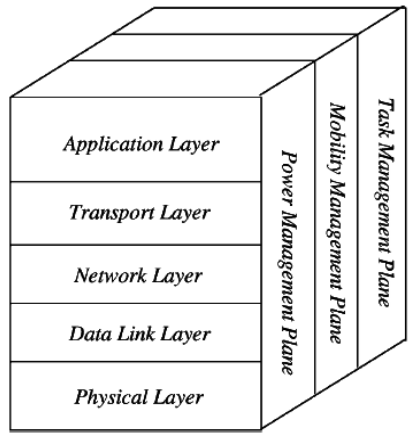

Figure 3: WSN Protocol Stack

In addition, the Power, Mobility, and Task Management planes monitor the power, movement, and task distribution among the sensor nodes. These planes help the sensor nodes coordinate the sensing task and lower the overall energy consumption [10]

## 7. SECURITY CHALLENGES ON WSN

Attacks on Wireless Sensor Network is subdivided into two categories; Passive and Active attacks.

### 7.1 Passive Attack

The conscious and unconscious violation of data privacy by listening to communication channels by an unauthorized person or attacker without actively physically tampering with the traversing data and medium of or channel of communication is termed a Passive Attack. It is an attack that particularly targets the privacy of information. These kinds of attack come in the following forms;

    a) ***Monitor and Eavesdropping***: This is the most common attack on privacy. By prying into a supposed confidential data, the attackers could easily discover the communication contents. When the traffic conveys the control information about the sensor network configuration, which contains potentially more detailed information than accessible through the location server, the eavesdropping can act effectively against the privacy protection.

    b) ***Traffic Analysis:*** Even when the messages transferred are encrypted, it still leaves a high possibility analysis of the communication patterns. Sensor activities can potentially reveal enough information to enable an attacker to cause malicious harm to the sensor network.

    c) ***Camouflage Adversaries:*** One can insert their node or compromise the nodes to hide in the sensor network. After that these nodes can copy as a normal node to attract the packets, then misroute the packets, conducting the privacy analysis [11]

**7.2 Active Attack**

As the name sounds "Active", which denotes some sort of physical activity. This attack involves a physical access to either the data, the channel or medium of communication or both. The attacker listens to and modifies the channel with or without the knowledge of the data and communication channel owner. These kinds of attack come in the following forms:

1) Routing Attacks
2) Denial of Service Attacks
3) Node Subversion
4) Node Malfunction
5) Node Outage
6) Physical Attacks
7) Message Corruption
8) False Node
9) Node Replication Attacks
10) Passive Information Gathering

Below is a detailed explanation of each of these attacks:

**1. Routing Attacks**

The attacks which act on the network layer are called routing attacks. The following are the attacks that happen while the message is been routed through the network.

*a. Spoofed, altered and replayed routing information*

  i. An unprotected ad-hoc routing is vulnerable to these types of attacks, as every node acts as a router, and can therefore directly affect routing information.
  ii. Create routing loops
  iii. Extend or shorten service routes
  iv. Generate false error messages
  v. Increase end-to-end latency

*b. Selective Forwarding*

A malicious node can selectively drop only certain packets. Especially effective if combined with an attack that gathers much traffic via the node. In sensor networks, it is assumed that nodes faithfully forward received messages. But some compromised node might refuse to forward packets, however, neighbors might start using another route [11].

*c. Sinkhole Attack*

Attracting traffic to a specific node in called sinkhole attack. In this attack, the attacker's goal is to attract nearly all the traffic from a particular area through a compromised node. Sinkhole attacks typically work by making a compromised node look especially attractive to surrounding nodes [12].

*d. Sybil Attacks*

A single node duplicates itself and presented in the multiple locations. The Sybil attack targets fault tolerant schemes such as distributed storage, multipath routing, and topology maintenance. In a Sybil attack, a single node presents multiple identities to other nodes in the network. Authentication and encryption techniques can prevent an outsider to launch a Sybil attack on the sensor network [11].

*e. Wormholes Attacks*

In the wormhole attack, an attacker records packets (or bits) at one location in the network, tunnels them to another location and retransmits them into the network [12].

*f. HELLO flood attacks*

An attacker sends or replays a routing protocol's HELLO packets from one node to another with more energy. This attack uses HELLO packets as a weapon to convince the sensors in WSN. In this type of attack, an attacker with a high radio transmission range and processing power sends HELLO packets to a number of sensor nodes that are isolated in a large area within a WSN. The sensors are thus influenced that the attacker is their neighbor. As a result, while sending the information to the base station, the victim nodes try to go through the attacker as they know that it is their neighbor and are ultimately spoofed by the attacker [12].

**2. Denial of Service Denial of Service (DoS)**

DoS happens by the unintentional failure of nodes caused by the malicious action or faulty nodes. DoS attack is meant not only for the attacker's attempt to subvert, disrupt, or destroy a network but also for any event that diminishes a network's capability to provide a service. In wireless sensor networks, several types of DoS attacks in different layers might be performed. At physical layer the DoS attacks could be jamming and tampering, at the Data Link layer, collision, exhaustion, and unfairness, at the Network layer, neglect and greed, homing, misdirection, black holes and at Transport layer this attack could be performed by malicious flooding and de-synchronization. The mechanisms to prevent DoS attacks include payment for network resources, push-back, strong authentication and identification of traffic [9].

**3. Node Subversion**

Capture of a node may reveal its information including disclosure of cryptographic keys and thus compromise the whole sensor network. A particular sensor might be captured, and information (key) stored on it might be obtained by an attacker [11].

**4. Node Malfunction**

A malfunctioning node will generate inaccurate data that could expose the integrity of sensor network especially if it is a data-aggregating node such as a Cluster Head or Sink [11].

## 5. Node Outage

Node outage is the situation that occurs when a node stops its function. In the case where a Cluster Head stops functioning, the sensor network protocols should be robust enough to mitigate the effects of node outages by providing an alternate route [11]

## 6. Physical Attacks

Sensor networks typically operate in hostile outdoor environments. In such environments, the small form factor of the sensors, coupled with the unattended and distributed nature of their deployment make them highly susceptible to physical attacks, i.e., threats due to physical node destructions. Unlike many other attacks mentioned above, physical attacks destroy sensors permanently, so the losses are irreversible. For instance, attackers can extract cryptographic secrets, tamper with the associated circuitry, modify programming in the sensors, or replace them with malicious sensors under the control of the attacker [13].

## 7. Message Corruption

Any modification of the content of a message by an attacker compromises its integrity.

## 8. False Node

A false node involves the addition of a node by an attacker and causes the injection of malicious data. An intruder might add a node to the system that feeds false data or prevents the passage of true data. Insertion of the malicious node is one of the most dangerous attacks that can occur. Malicious code injected into the network could spread to all nodes, potentially destroying the whole network, or even worse, taking over the network on behalf of an attacker [11].

## 9. Node Replication Attacks

Conceptually, a node replication attack is quite simple; an attacker seeks to add a node to an existing sensor network by copying the node ID of an existing sensor node. A node replicated in this approach can severely disrupt a sensor network's performance. Packets can be corrupted or even misrouted. This can result in a disconnected network, false sensor readings, etc. If an attacker can gain physical access to the entire network he can copy cryptographic keys to the replicated sensor nodes. By inserting the replicated nodes at specific network points, the attacker could easily manipulate a specific segment of the network, perhaps by disconnecting it all together [11].

## 10. Passive Information

Gathering an attacker with powerful resources can collect information from the sensor networks if it is not encrypted. An intruder with an appropriately powerful receiver and well-designed antenna can easily pick off the data stream. Interception of the messages containing the physical locations of sensor nodes allows an attacker to locate the nodes and destroy them. Besides the locations of sensor nodes, an attacker can observe the application-specific content of messages including message IDs, timestamps, and other fields. To minimize the threats of passive information gathering, strong encryption techniques need to be used [12].

## 8. WSN SECURITY ATTACKS AND MITIGATION TECHNIQUES: THE WSN PROTOCOL STACK APPROACH

The table below presents the WSN Network Protocol Stack and the types of possible attacks that can be carried out on each layer as well as the mitigation techniques on these attacks.

| WSN Network Protocol Stack Layers | Security Attacks | Mitigation Techniques |
|---|---|---|
| Application | Attacks on reliability and Clone attack: Clock skewing, Selective message forwarding, Data aggregation distortion | i. Unique pairwise keys and cryptographic approach. ii. Authentication can be used to protect any data integrity iii. Encryption is an effective approach for data confidentiality protection |
| Transportation | Flooding | Client Puzzles |
| | Desynchronization | Authentication |
| Network Layer | Spoofed, altered or replayed routing information | Authentication, Monitoring |
| | Selective forwarding | Probing, Redundancy |
| | Sinkhole | Monitoring, Redundancy, Authentication |
| | Sybil | Probing, Authentication |
| | Wormholes | Authentication, Packet leashes by using geographic and temporal information |
| | Hello flood | Verify the bidirectional link, Authentication |
| | Acknowledgment spoofing Flooding | Authentication |

| WSN Network Protocol Stack Layers | Security Attacks | Mitigation Techniques |
|---|---|---|
| Data Link | Collision | Error-correcting code |
| | Exhaustion | Rate limitation |
| Physical | Jamming | Lower duty cycle, Priority messages, Spread-spectrum techniques |
| | Tampering | Tamper-proofing, Hiding |

Table 1: WSN Security Attacks and Mitigation Techniques

## 9. CONCLUSION

As Wireless Sensor Network gradually becomes popular no doubt Attackers will shift their attention from the traditional Network to that of the Wireless Sensor Network because of the sensitive data that traverses WSN. As it has already begun, in the near future, homes will go smart, factories, agriculture and every aspect of our lives will be automated and controlled by smart sensors and hand-held devices respectively. Conscious efforts have to be made now to secure WSN which is the backbone of the Smart environment. Owing to the sensitive nature of data or information traversing WSN Attackers will use either passive or active attack mechanism to compromise the integrity of data or disrupt the smooth operations of the network. Unfortunately, because it is very difficult to stop Attackers from launching an attack, one can only find ways to mitigate or minimize the severity of an attack in other to keep the network operational while fixing the identified problem. This is an open challenge for Researchers to look for more effective ways to offer mitigating solutions to various kinds of attacks as they emerge. In this Paper, efforts have been made to shed light on the various security threats to WSN as well as discuss current mitigating techniques.